# Internal-wave convection and shear near the top of a deep equatorial seamount

**by Hans van Haren**

Royal Netherlands Institute for Sea Research (NIOZ), P.O. Box 59, 1790 AB Den Burg, the Netherlands.
e-mail: hans.van.haren@nioz.nl

*Abstract*--The near-equatorial ocean experiences particular dynamics because the Coriolis force is weak. One modelled effect of these dynamics is strong reduction of turbulent mixing in the ocean interior. Unknowns are effects on internal wave breaking and associated turbulent mixing above steeply sloping topography. In this paper, high-resolution temperature observations are analyzed from sensors that were moored near the top of a deep Ceará Basin seamount for one week. A vertical string held sensors between 0.4 and 56.4 meters above the seafloor. The observations show common semidiurnal-periodic internal wave breaking, with tidal- and 56-m mean turbulence values that are not significantly different from those observed near the top of 1000-m shallower mid-latitude Great Meteor Seamount, despite the twice lower vertical density stratification. Profiles of 6-day mean turbulence values yield vertically uniform values except for a small decrease in the lower 2 m above the seafloor. The lower 2-m show a distinct departure from turbulent inertial subrange in temperature variance spectra. In 10-m higher-up, spectral slopes indicate dominant turbulent convection with reduced flow and turbulence, except when a primary tidal bore is present. Further-up than 15 m, shear dominates (stratified) turbulence. The lack of Coriolis force is not found to be important for internal wave-induced turbulence above steeply sloping topography, except that Kelvin-Helmholtz instabilities seem somewhat less chaotic and more organized roll-up than at mid-latitudes.

*1. Introduction*

The idea that the ocean remains vertically stably stratified via sufficient turbulent mixing (Munk, 1966; Munk & Wunsch, 1998; Wunsch & Ferrari, 2004) provided the suggestion that most mixing occurs near sloping seafloors (e.g., Armi 1978; Armi 1979). Most turbulent mixing near underwater topography seems to be invoked by breaking internal waves (Wunsch, 1970; Eriksen, 1982; Thorpe, 1987; Sarkar & Scotti, 2017). Other agents, such as friction by large-scale flows over flat seafloors or flows over topography lead to turbulent mixing rates that are at least one order of magnitude smaller (e.g., Nikurashin et al., 2014). Bio-mixing triggered by massive migrations such as diurnal zooplankton motions basically lead to direct conversion of mechanical energy into heat without turbulent overturning, due to the small length-scales (Visser, 2007). However, also not all internal wave breaking leads to sufficiently large turbulent mixing. Variations in turbulence values have been observed over one-two orders of magnitude across horizontal distances $O(10$ km$)$ above underwater topography (Nash et al., 2007; van Haren et al., 2015). Most turbulent mixing occurs over steep slopes, not only related with canyons, to such extent that it appears to be sufficient to maintain the global ocean stratification.

In contrast with steady flows inducing the type of shear-turbulence via friction above a flat seafloor, and in contrast with internal-wave shear in the ocean interior, waves breaking at a slope may induce the type of convection-turbulence when they overturn. Although both types of turbulence occur in exchange with each other, for example secondary shear occurring in thin layers adjacent to primary convection-plumes (Li & Li, 2006), the dominance of one of the two may be important for the extent of the mixing. As a result, in the ocean interior shear is dominated by near-inertial motions because of their small vertical scales (e.g., LeBlond & Mysak, 1978). Thereby, Earth rotational (Coriolis) effects come into play. Rotational effects are also known to guide the plumes in chimneys of free convection (Julien et al., 1996).

It is unclear whether and how rotational effects play a role in internal wave breaking above underwater topography. To investigate in this paper, observational data are presented from high-resolution temperature sensors that were moored in a small canyon near the top of a deep equatorial seamount. In this area, rotational Coriolis effects are expected to collapse (e.g., Veronis, 1963). The sensors were mounted to a bottom-lander mooring and reached to within 0.4 m from the seafloor, in a similar fashion as has been used near the summit of mid-latitude Great Meteor Seamount (GMS) (van Haren and Gostiaux, 2012). Some comparison will be made between the near-equatorial and the GMS data, including spectral distinction between shear- and convection-turbulence as a function of distance above the seafloor.

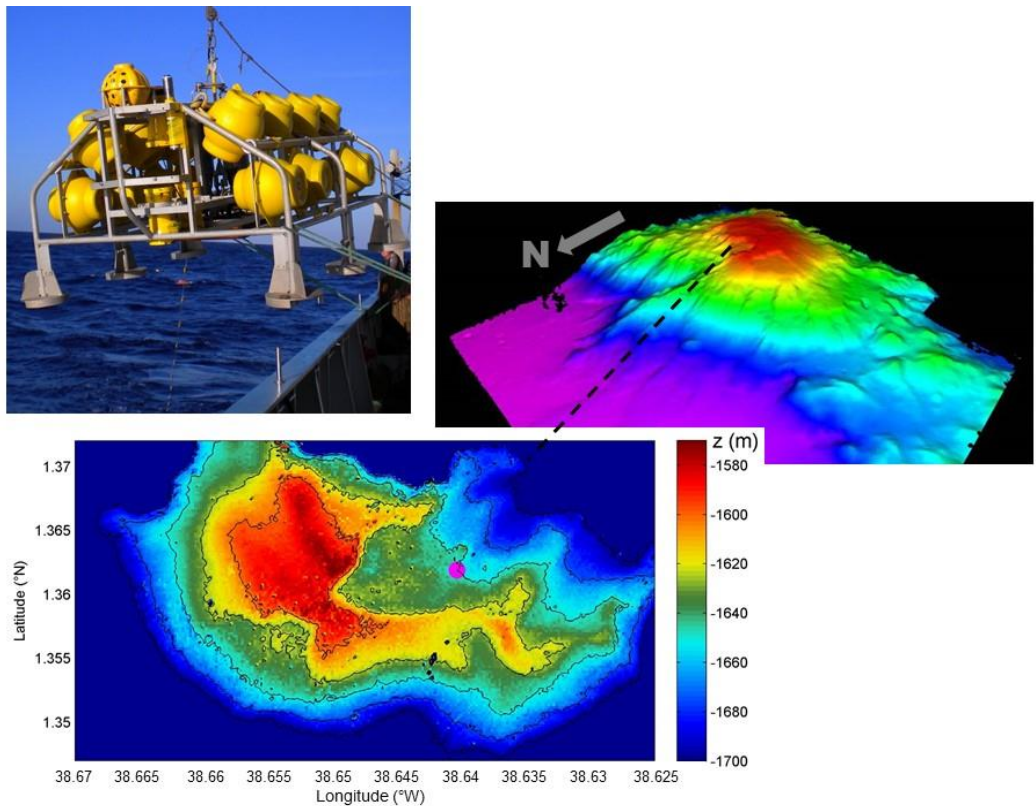

Figure 1

Bottom lander frame with thin line holding temperature sensors to the orange top-buoyancy in the far distance at the sea surface, just prior to deployment, together with Multibeam map of unnamed 2500-m tall near-equatorial deep seamount in Ceará Basin NW-Atlantic Ocean. The seamount's summit is at about z = -1570 m and surrounding waters reach z < -4000 m. The summit detail, facing north with different colour scale, shows the location of T-string mooring (purple dot)

## 2. Technical details

*2.1. Site and instrumentation*

To study near-equatorial internal wave-induced turbulence dynamics above sloping topography, a 75-m tall taut-wire mooring was located at 01° 21.714′N, 038° 38.421′W, H = 1648 m water depth. The mooring was approximately in the thalweg of a small canyon near the 1570-m deep top of an unnamed seamount in Ceará Basin, NW-Atlantic Ocean (Fig. 1). The mooring was underwater for 6.7 days between 23-30 June, days 173-180, in 2009.

The local seafloor slope was about $\gamma = 0.09$ (5°) computed over 1 km horizontal distance, about half that value (less steep slope) along the thalweg higher up, and larger (steeper) values deeper down. The local slope is supercritical for linear propagating semidiurnal lunar $M_2$ internal tidal waves, $\gamma > \beta_{M2} = \sin^{-1}((\omega^2-f^2)^{1/2}/(N^2-f^2)^{1/2}) = 0.06$ (3.5°), e.g., LeBlond & Mysak (1978), in which $\omega$ denotes the internal wave frequency ($\omega_{M2} = 1.405 \times 10^{-4}$ s$^{-1}$), $f = 3.466 \times 10^{-6}$ s$^{-1}$ the weak near-equatorial planetary inertial frequency (Coriolis parameter), and $N = 2.3 \times 10^{-3}$ s$^{-1}$ the 100-m large-scale mean local buoyancy frequency (Fig. 2). The local horizontal Coriolis parameter, which is relevant for near-equatorial dynamics at latitudes $|\varphi| < 2°$ (e.g., Veronis, 1963; Munk & Moore, 1968), amounts $f_h = 1.458 \times 10^{-4}$ s$^{-1}$.

The mooring was held tautly upright by a single sub-surface 'top'-buoy at about h = 75 m above the seafloor. The top-buoy provided about 1.7-kN net buoyancy to the entire mooring assembly. Under maximum 0.3 m s$^{-1}$ current amplitudes, the low-drag mooring did not deflect from the vertical by more than 2°, i.e., the top-buoy moved <6 m horizontally and <0.04 m vertically, as inferred from tilt and pressure sensors.

The top-buoy was attached via a nylon-coated 0.005-m diameter steel cable to a bottom-lander frame (Fig. 1) holding two acoustic releases, an anchor weight and an upward looking 300-kHz TeleDyne/RDI four-beam acoustic Doppler current profiler (ADCP). The ADCP sampled 80 1.0 m vertical bins between h = 6 and 85 m, at a rate of once per 2 s. Due to low amounts of acoustic scatterers in the local deep-ocean waters, the resolved vertical range was barely 40-50 m, depending on the phase of the tide.

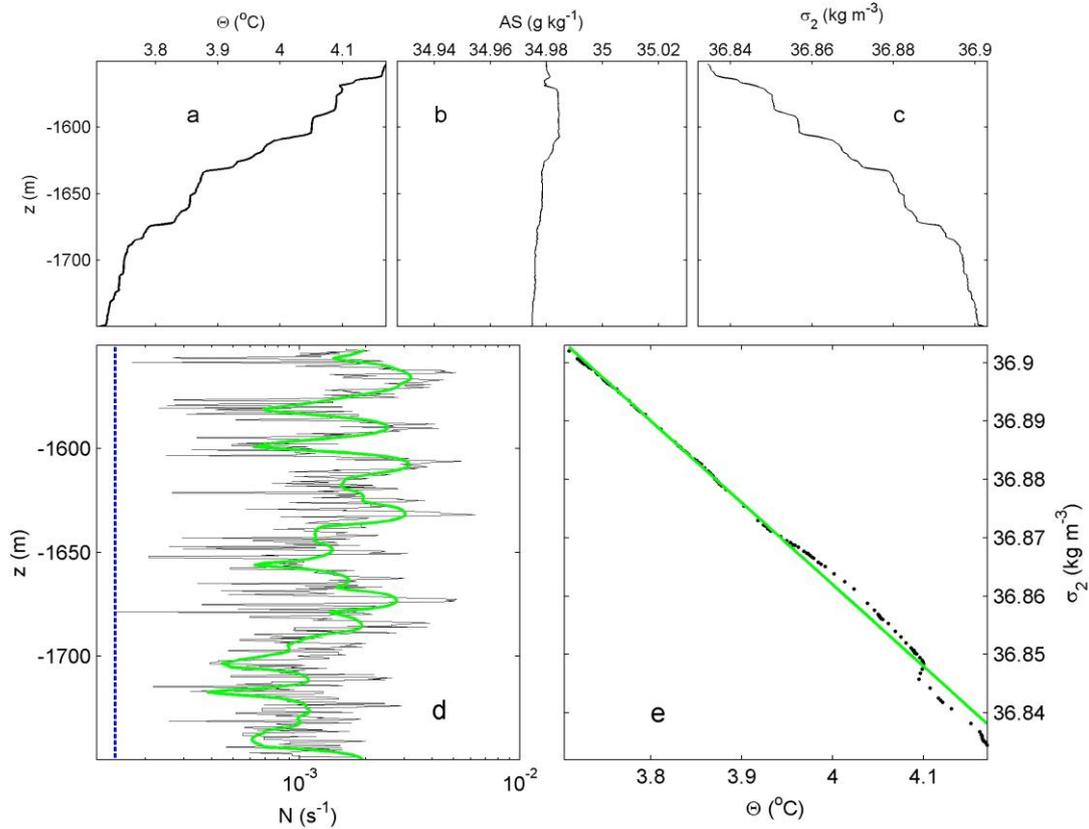

Figure 2

Shipborne CTD-observations over 200-m vertical range, obtained just to the north of the T-sensor mooring but in 2250 m water-depth two days after mooring deployment. (a) Conservative Temperature. (b) Absolute Salinity. The x-axis range approximately corresponds with that of a. in terms of density contributions. (c) Density anomaly referenced to 2000 dbar ($2\times10^7$ Pa). (d) Buoyancy frequency computed over 1 (black) and 10 dbar (green) intervals from the unsorted density profile in c. The blue dashed line indicates the approximate threshold of $1f_h = 1.458\times10^{-4}$ s$^{-1}$, see text. (e) Conservative Temperature-density anomaly relationship with linear fit in green, from the CTD-profile portions in a. and c.

A total of 55 'NIOZ3' high-resolution temperature T-sensors were taped to the mooring cable and 2 more to the bottom lander. The T-sensors were at 1.0 m intervals in the range of h = 0.4-56.4 m, and sampled at a rate of once per 1 s.

NIOZ3 are self-contained T-sensors, with a sensor tip smaller than 1 mm, a response time of <0.5 s, an initial drift of about $1\times10^{-3}$ °C mo$^{-1}$ after aging of the thermistor electronics, a noise level of <$1\times10^{-4}$ °C, and a precision (relative accuracy) better than $5\times10^{-4}$ °C after drift-correction for standard programmed measuring of the Wien Bridge oscillator for the duration of 0.12 s per sample (van Haren et al., 2009). Every four hours, all sensors are synchronized via induction to a single standard clock, so that the entire 56-m vertical range is sampled in less than 0.02 s. Of the 57 T-sensors, 13 showed calibration, noise, battery, or other electronic problems. The data of these sensors are replaced by linear interpolation between neighboring sensors.

During post-processing, the in situ CTD-calibrated temperature data are bias-corrected to a smooth low-order polynomial mean profile and converted into 'Conservative' (~potential) Temperature data $\Theta$ (IOC, SCOR, IAPSO, 2010). Although NIOZ3 T-sensors have relatively low noise levels, they show relatively large bias mainly due to drift compared to later versions. As a result, several sensors did not pass the bias-correction. These are included in the data-interpolation above. Corrected moored T-sensor data can be used to compute turbulence values, as elaborated below. Interpolation low-biases turbulence values, which data are excluded from vertically averaged values. However, interpolation because of poor bias-correction is not necessary for spectral analysis, as bias is a very low-frequency phenomenon, so that uncorrected data can be used.

*2.2. Computing turbulence from moored T-sensor data*

The post-processed moored T-sensor data act as a tracer for variations in density anomaly $\sigma_{1.6}$ following the relation,

$$\delta\sigma_{1.6} = \alpha\delta\Theta, \alpha = -0.13\pm0.005 \text{ kg m}^{-3} \text{ °C}^{-1}, \tag{1}$$

where subscript 1.6 indicates a pressure reference of 1600 dbar (1 dbar = $10^4$ Pa). This temperature-density relation is established from data between -1750 < z < -1550 m of a shipborne Conductivity-Temperature-Depth (CTD)-profile (Fig. 2) obtained about 3 km from the mooring site. The CTD-data are also used for referencing the moored T-sensors during bias-correction.

Given the reasonably tight relationship (1), the number of moored T-sensors and their spacing of 1.0 m, in combination with their low noise level and precision, allows for accurately calculating values of turbulent kinetic energy dissipation rate ε and eddy diffusivity $K_z$ via the reordering of unstable overturns making vertical density profiles statically stable (Thorpe, 1977). These overturns follow reordering every 1 s the 56-m high potential density (Conservative Temperature) profile $\sigma_{1.6}(z)$, which may contain unstable inversions, into a stable monotonic profile $\sigma_{1.6}(z_s)$ without inversions.

After comparing observed and reordered profiles, displacements d = min(|z-$z_s$|)·sgn(z-$z_s$) are calculated necessary to generate the reordered stable profile. Test-thresholds are applied to disregard apparent displacements associated with instrumental noise and post-calibration errors (Galbraith & Kelley, 1996). Such a test-threshold is very low $<5\times10^{-4}$ °C for originally sampled non-spiked moored NIOZ T-sensor data. Otherwise, identical methodology is used as proposed for CTD-data in (Thorpe, 1977) and applied for processing mooring data by, e.g., (Levine & Boyd 2006; Aucan et al., 2006; Nash et al., 2007), see (van Haren & Gostiaux, 2012) for details. It includes a constant mixing efficiency of 0.2 (Osborn, 1980; Oakey, 1982), an Ozmidov $L_O = (\varepsilon/N^3)^{0.5}$ (Dougherty, 1961; Ozmidov, 1965) -- root-mean-square (rms) overturn scale $d_{rms} = (1/n\Sigma d^2)^{0.5} = L_T$ ratio of $L_O/L_T = 0.8$ (Dillon, 1982) over many-n samples, and the computation of local small-scale buoyancy frequency $N_s(t,z)$ from the reordered stable density (temperature) profiles (Thorpe, 1977; Dillon, 1982).

Although individual values of mixing efficiency vary over at least an order of magnitude (Oakey, 1982; Dillon, 1982), moored T-sensor data allow for sufficient averaging to warrant the use of constant mixing efficiency in the computation of turbulence values. When sufficient averaging over time and/or depth is applied, various types, conditions and stages of turbulence are included, each with different efficiency (Mater et al., 2015; Cyr & van Haren, 2016). Large-scale overall mean N is computed as rms($N_s$), a linear operation as $N^2 \propto d\sigma_{1.6}/dz$. Then, using d rather than $d_{rms}$ as explained below,

$\varepsilon = 0.64d^2N_s^3$, (2)

$K_z = 0.2\varepsilon N_s^{-2}$. (3)

In (2), and thereby (3), individual d replace their rms-value across a single overturn as originally proposed by Thorpe (1977). The reason is that individual overturns cannot easily be distinguished, because overturns are found at various scales with small ones overprinting larger overturns, as one expects from turbulence. This procedure provides high-resolution time-vertical images of turbulence values, for qualitative studies. Subsequently for quantitative studies, 'mean' turbulence values are calculated by arithmetic averaging over the vertical <...> or over time [..], or both, which is possible using moored high-resolution T-sensors. This ensures the sufficient averaging required to use the above mean constant values (e.g., Osborn, 1980; Oakey, 1982; Mater et al., 2015; Gregg et al., 2018).

Using similar moored T-sensor data from GMS, van Haren & Gostiaux (2012) found turbulence values to within a factor of three like those inferred from ship-borne CTD/LADCP profiler data using a shear/strain parameterisation near the seafloor. Their values compare well with microstructure profiler-based estimates in similar sloping topography areas by Klymak et al. (2008). Comparison between calculated turbulence values using shear measurements and using overturning scales with $L_O/d_{rms} = 0.8$ from areas with mixtures of turbulence development above sloping topography led to consistent results (Nash et al., 2007), after suitable averaging over the buoyancy scales. Thus, from the argumentation above and the reasoning in Mater et al. (2015), internal wave breaking is unlikely to bias turbulence values computed from overturning scales by more than a factor of two to three, provided some suitable time-space averaging is done instead of considering single profiles. This is within the range of error, also of general ocean turbulence measurements using shipborne microstructure profilers (Oakey, 1982).

## 3. Observations

*3.1. Overview*

The vertical profiles of shipborne CTD data demonstrate generally steady decreases of temperature and salinity with depth, so that temperature is the dominant contributor to stable density variations with depth (Fig. 2). Occasionally, some inversion is seen in the salinity data only, but this is attributed to the lesser mixing at the corresponding height above the seafloor several kilometers away from the seamount where the CTD-profile was obtained. Near the steeply sloping seafloor around 2300 m water depth where the CTD was lowered, the temperature-salinity and thus temperature-density relationship is non-ambiguous and very tight (Fig. 11 in Appendix A). This CTD-observation near the local seafloor is portable to the site of the moored instrumentation above the slope higher-up. While the vertical hydrographic profiles do show some stepping, variations in vertical gradients of density, and small inversions, the stratification is non-zero as small-scale $N_s > f_h$. A minimum $N_{min} = f_h$ marks a threshold under nonlinear stability at mid-latitudes (van Haren, 2008), and is apparently extendable to the near-equatorial band $|\varphi| < 2°$.

Time series of temperature show the familiar characteristics of semidiurnal internal tidal waves near a steep seafloor slope (Fig. 3a), with slowly increasing temperature with multiple small-scale variations during the warming phase and rapidly decreasing temperature during the cooling phase. It also shows familiar considerable variations in timing and amplitude between the tidal periods, and strong phase differences with records observed only 50 m higher up. In addition, the present near-seafloor observations demonstrate a visible decrease in small-scale temperature variability over only 2-m vertical difference, with least variability at h = 0.4 m.

The relatively poorly resolved and noisy ADCP data demonstrate the current speeds being <0.3 m s$^{-1}$ (Fig. 3b), with larger speeds higher up and smallest speeds between days 175 and 178 when near-seafloor temperature showed largest semidiurnal variation. The current speeds' semidiurnal periodicity points at non-circular current ellipses, of which the major axis commonly is directed along the thalweg.

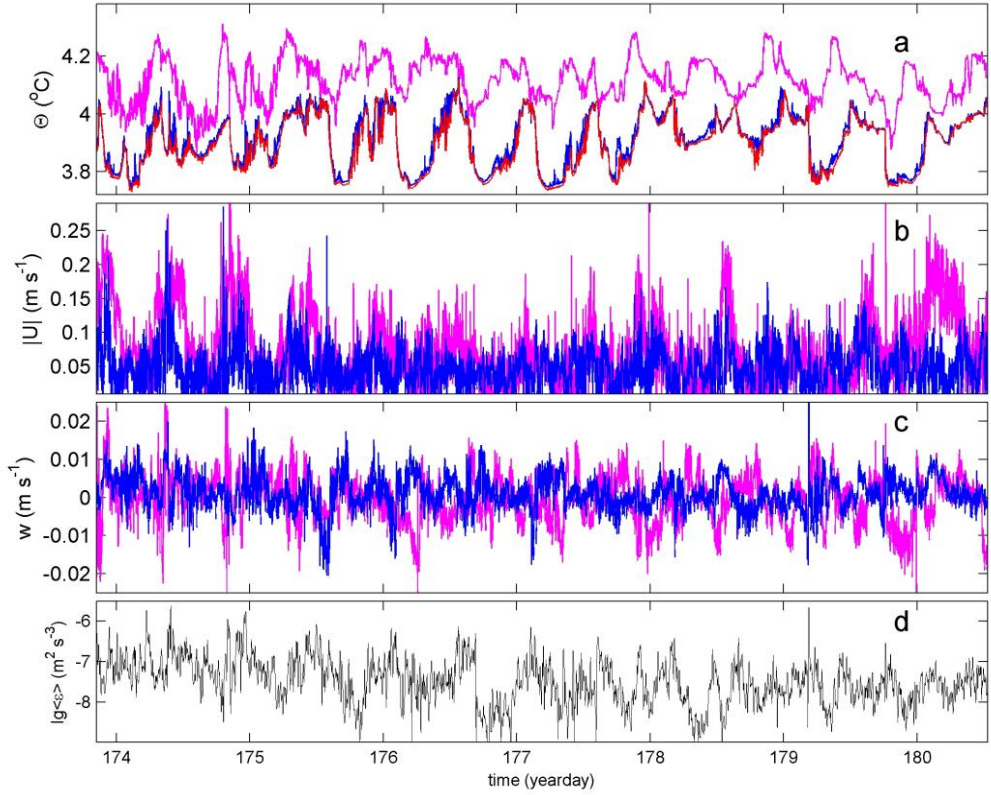

Figure 3

6.7-day time series of mooring data. (a) Conservative Temperature from T-sensors at h = 0.4 (red), 2.4 (blue) and 55.4 m (magenta). (b) Three-minute smoothed current amplitudes from ADCP at h = 7 (blue) and 27 m (magenta). (c) Three-minute smoothed vertical current component from ADCP at h = 10 (blue) and 15 m (magenta). (d) Six-minute smoothed logarithm of turbulent kinetic energy dissipation rate averaged over the vertical T-sensor range for each profile

Vertical velocities (Fig. 3c) show peaks >0.02 m s$^{-1}$ occurring at semidiurnal periodicities but associated with high-frequency internal waves near the buoyancy frequency, e.g., on days 174-175, or frontal bores, e.g., on day 179.19. The latter bore associates with a peak in turbulent kinetic energy dissipation rate (Fig. 3d). The vertically averaged dissipation rate shows coarsely a semidiurnal periodicity, and peaks that do occur at several steep temperature drops, but not at all of them. This indicates a high variability in frontal-bore turbulence reaching the seafloor, or not, and semidiurnal tidal wave breaking in general. To compare temperature and turbulence distributions, non-averaged values are qualitatively considered in the overview of Fig. 4.

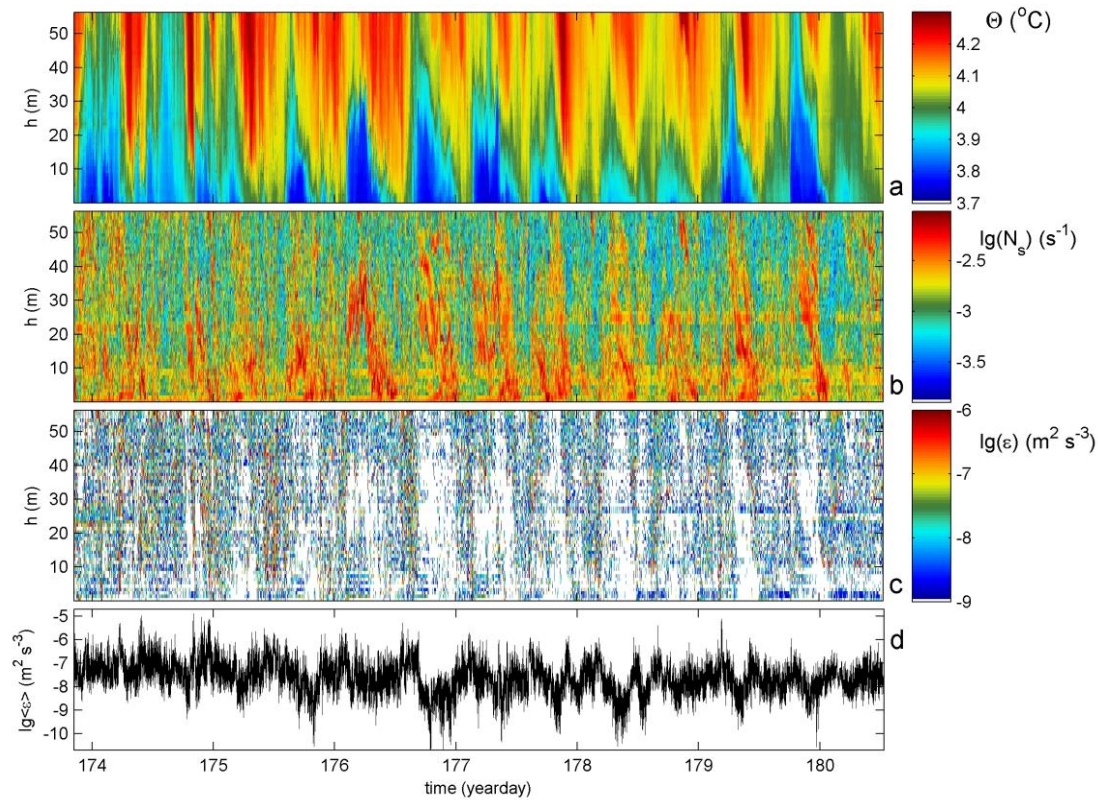

Figure 4

6.7-day 56-m-tall, moored T-sensor data and calculated turbulence values. In a.-c., the horizontal axis is at the level of the local seafloor. (a) Conservative Temperature. (b) Logarithm of small-scale buoyancy frequency calculated from reordered profiles of a. (c) Logarithm of non-averaged turbulent kinetic energy dissipation rate. White values are below threshold. (d) Logarithm of turbulent kinetic energy dissipation rate averaged over the vertical T-sensor range for each profile

The 6.7-day time-depth series of T-sensor data shows that the full vertical extent of the semidiurnal internal tides is not resolved by the 56-m vertical range (Fig. 4a). The non-linear sawtooth-like tidal forms are superseded with high-frequency internal waves. They combine, in phase-locked fashion, with the semidiurnal waves to the nonlinear forms. The small-scale stratification as computed from the stable reordered temperature profiles is generally found closer to the seafloor than above (Fig. 4b). Most intense near-seafloor stratification often occurs towards the end of the warming phase and is being swept up to h = 20-30 m during the upslope moving cooling phase. The apparent banding in moderate stratification is partially artificial due to unresolvable bias of NIOZ3-sensors, which little affects (peaks in) turbulence and stratification values and is partially real (in the lower h < 10 m).

Although the computation of turbulence values from reordering of instabilities forces one to consider vertical or time averages over entire overturns, it is instructive to consider the pattern of non-averaged values (Fig. 4c). This image of non-averaged turbulent kinetic energy dissipation rate values demonstrates extremely few short-lived moments when turbulence reaches the seafloor. These moments occur during frontal bore passages with intense turbulent overturning possibly affecting sediment resuspension. Examples are on days 176.64 and 179.19 for upslope moving bores, while the peak on day 174.85 is for a downslope moving frontal breaker. The former are backwards breaking waves, the latter a forward breaker.

While some of these overturns are large having >10-m vertical extent, as will be demonstrated below, generally larger overturns are observed just away from the seafloor, depressing stratification to within a few meters from the seafloor, mostly during the warming phase of the tide. This leads to rapid restratification that is typical for internal wave breaking over supercritical slopes (Winters, 2015), and makes the turbulent mixing quite efficient. In between the breaking waves, relatively quiescent periods occur with two orders of magnitude lower turbulence values (Fig. 4d), which are still at least one order of magnitude larger than values found in the ocean interior (e.g., Gregg, 1989).

The 6.7-day mean turbulent kinetic energy dissipation rate is $[<\varepsilon>] = 9\pm6\times10^{-8}$ m$^2$ s$^{-3}$, the mean eddy diffusivity $[<K_z>] = 3\pm1.5\times10^{-3}$ m$^2$ s$^{-1}$ under mean buoyancy frequency N = $[<N_s>] = 2.3\pm0.3\times10^{-3}$ s$^{-1}$. These mean turbulence values are not significantly smaller than found by van Haren & Gostiaux (2012) above a semidiurnal-internal-wave supercritical slope at 549 m near the top of GMS in about twice larger stratification. Like above GMS, half of the mean turbulence values are reached during 6% of the time when values are more than half an order of magnitude larger than the mean value, see Fig. 4d.

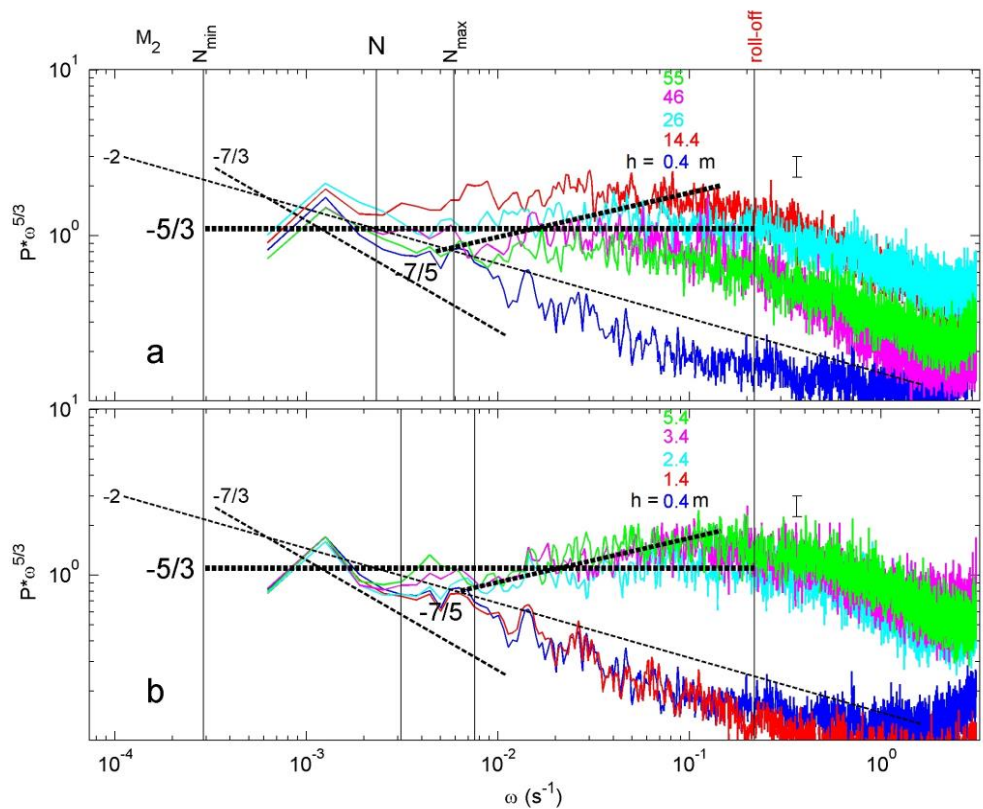

Figure 5

Moderately smoothed (about 30 degrees of freedom) temperature variance frequency ($\omega$) spectra focusing on the turbulence range and scaled with the inertial subrange slope $\omega^{-5/3}$, for 6.7-day data from vertical levels as indicated by their distance h from the seafloor (in meters). Several frequencies (see text) are indicated with vertical lines, except for semidiurnal tidal $M_2$. Four spectral slopes are indicated by dashed black lines and by their exponent-value (for unscaled log-log plots). (a) Coarsely covering the entire range of observations. (b) Focusing on the lower 5 T-sensors providing good data

*3.2. Mean spectra*

To investigate variations in the turbulence range of temperature data as a function of distance from the sloping seafloor, the 6.7-day mean spectra from nine T-sensors are computed (Fig. 5). For clarity of focus on the turbulence range, the spectra are scaled with the frequency ($\omega$) slope $\omega^{-5/3}$ of most common shear-induced inertial subrange. The spectra are plotted in two sets of five, with the lowest T-sensor functioning as a reference between both panels together with several spectral slopes on the log-log plots. In Fig. 5a, the entire 56-m range is covered. In Fig. 5b, spectra from all well-functioning T-sensors in the lower h ≤ 5.4 m are presented.

The plotted reference slopes are indicated by their exponent-value in unscaled plots: $\omega^{-2}$ for spectra of general internal waves outside inertial, tidal (harmonic) and buoyancy frequencies (Garrett & Munk, 1972) and also for fine-structure contamination (Phillips, 1971; Reid, 1971), $\omega^{-5/3}$ for the inertial subrange of turbulence predominantly induced by shear and representing passive scalars (Tennekes & Lumley, 1972; Warhaft, 2000), and $\omega^{-7/5}$ as an example of significant deviation from inertial subrange for turbulence predominantly induced by buoyancy-driven convection and representing active scalars (Bolgiano, 1959; Pawar & Arakeri, 2016). The steepest indicated slope of $\omega^{-7/3}$ is a fit to the high-frequency internal wave and low-frequency turbulence part of the buoyancy subrange of the spectrum. It represents intermittency as, e.g., observed in occurrence of sea-surface white-capping (Malila et al., 2022). Local effects of intermittency attributed to coherent structures resulted in a dip in atmospheric velocity variance at the low end of the inertial subrange (Szilagy et al., 1996).

Vertical black lines indicate frequencies delimiting predominant internal waves at various scales and turbulence. Instrumental noise is found basically only at the far high extreme $\omega > 2$ s$^{-1}$, close to the Nyquist frequency. Internal waves associate with $\omega \in$ [f, (N$_{min}$) N (N$_{max}$)], of which the range [f, N] is commonly thought to be dominated by undamped linear and unsaturated internal waves and $\omega > N$ to consist of damped nonlinear and saturated internal waves (Weinstock, 1978; Munk, 1981). Turbulence is found in various identities in the range $\omega \in$ [ (N$_{min}$)…(N$_{max}$), roll-off] in which slopes are generally less steep than $\omega^{-2}$, except for (parts of) $\omega < N$. Motions at $\omega >$ roll-off have a slope between $\omega^{-2}$ and $\omega^{-7/3}$, before reaching instrumental white noise levels. Due to the shortness of records, spectra are moderately smoothed and errors thus relatively large.

In Fig. 5a, the poorly resolved internal wave band shows marginally increasing temperature variance away from the seafloor. For $\omega > N$, the discrepancy between different heights above the seafloor becomes significantly different, notably for $\omega > N_{max}$, which is the starting frequency of small temperature-variance differences, to the maximum discrepancy of about one order of magnitude, just before the roll-off frequency. The maximum discrepancy is found roughly around the Ozmidov frequency $\omega_O = U/L_O$. Using mean values, one finds $\omega_O = 0.05 \pm 0.02$ s$^{-1}$ but uncertainties are large, especially in U and precise matching of correct association of turbulent kinetic energy dissipation rate. Most of this temperature-variance spread is due to changes in the lower $h < 2$ m (Fig. 5b), as for $\omega > \omega_O$ and $h > 2.4$ m the spectra cluster together and become horizontal at about the same variance.

For $h > 2$ m, a relative dip of minimum (scaled) variance is observed between N and $N_{max}$. The poorly resolved spectral slope towards this minimum is between -2 and -7/3. At $\omega > N_{max}$, either the spectral slope follows -7/5 going up in the scaled plot (for $2 < h < 10$ m mean unscaled slope of $-1.40 \pm 0.10$ for [$N_{max}$, $3N_{max}$], displaying local differences in Fig. 5a, and -1.54±0.06 for [0.01 0.1]s$^{-1}$ displaying local differences notably in Fig. 5a), remains horizontal (for $10 < h < 55$ m mean unscaled slope of $-1.57 \pm 0.10$ for [$N_{max}$, $3N_{max}$] and $-1.73 \pm 0.04$ for [0.01 0.1]s$^{-1}$), or continues to slope like [-2, -7/3] before becoming horizontal (for $h < 2$ m mean unscaled slope of $-2.22 \pm 0.07$). The slope-change between spectra for $h = 14$ and 55 m is small, and their intermediate spectra around $h = 30$ m are mostly horizontal, although deviations towards convection-slope occur, between $N_{max} < \omega < 0.08$ s$^{-1}$ – roll-off.

The 'dip' has also been observed in spectra from towed temperature measurements and include a slope of -5/2 in the lower wavenumber range and -5/3 in the higher wavenumber range (Klymak & Moum, 2007), whereby deviations to -7/5-slope were not observed. This may be because those observations were not obtained very close, more than 50 m away, from sloping topography. In atmospheric sonic-velocity data at 5 m from the ground, the dip is visible around 10-m scales (Fig. 1 in Szilagy et al., 1996). In those data, dip-deepening had a slope <-2, while dip-filling can be recognized with a -7/5-slope between 1- and 10-m scales towards the resuming of the -5/3-slope at higher wavenumbers.

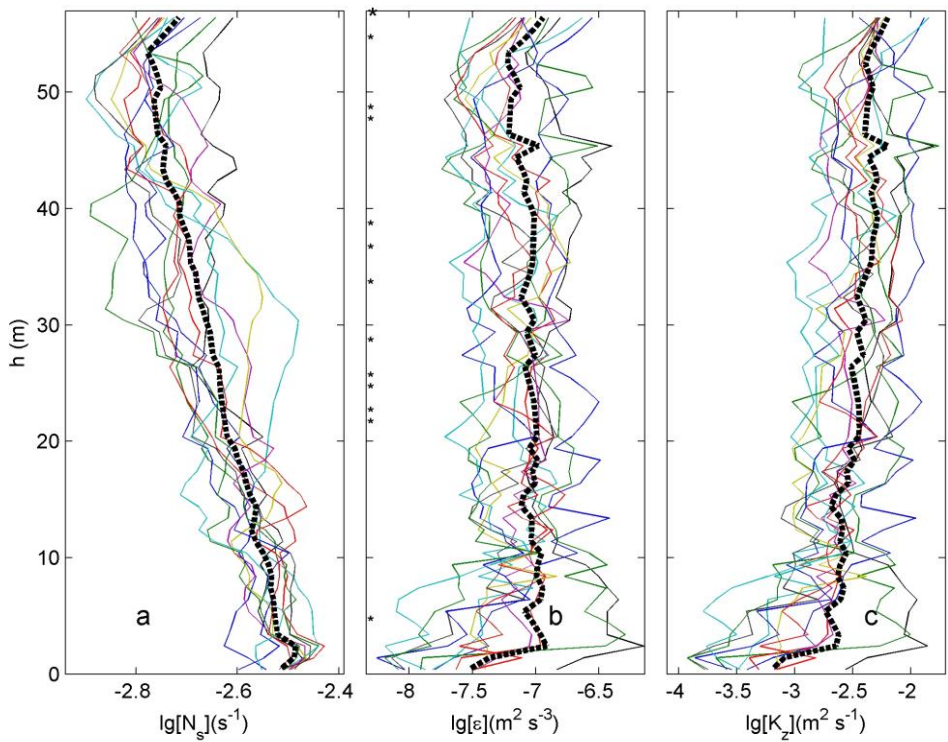

Figure 6

Profiles of 12 semidiurnal tidal-mean values of turbulence and stratification from the moored T-sensors (thin lines), and their 6-day mean values (thick black dashed). (a) Small-1-m-scale buoyancy frequency from the reordered, stable-only density profiles. (b) Turbulent kinetic energy dissipation rate. Values from data by malfunctioning sensors (stars to the left) are not considered and their mean values are interpolated between neighbouring sensor-depths. (c) Turbulent eddy diffusivity

Convection-turbulence spectral slopes have been observed in weakly stratified waters under breaking internal waves in Lake Garda (van Haren & Dijkstra, 2021). In the present observations it is seen between $2 < h < 25$ m, also predominantly enforced by breaking internal waves and unlikely due to general geothermal heating. The lower $h < 2$ m lacks small-scale turbulence motions in a rather smooth thin layer above the seafloor. It may be induced by the bottom-lander frame that may deviate flow around it, but it is noted that the two T-sensors were on the outside of the frame. The smooth near-seafloor layer has also been observed in the first few meters of general geothermal convective heating using cable-attached T-sensors in the deep Mediterranean (van Haren, 2023).

*3.3. Tidal-mean vertical profile variability*

To investigate the time-variation of tidally averaged turbulence profiles, data are averaged over the 12 consecutive semidiurnal periods and compared with their 6-day mean profile (Fig. 6). Outside the range of the bottom-lander frame, the first two data points above the seafloor, the overall mean buoyancy profile is seen to steadily decrease in value with increasing distance from the seafloor. Effects of the frame on the buoyancy frequency profiles of averaged stratification seem negligible and are considerably smaller than the variations between different semidiurnal mean profiles (Fig. 6a).

Although the tidally averaged values of the stratification yield distinctive vertical profiles with largest stratification reaching all the way to the sloping seafloor, as far as can be established, the associated profiles of turbulent kinetic energy dissipation rate (Fig. 6b) and eddy diffusivity (Fig. 6c) are less distinctive. In fact, the turbulent kinetic energy dissipation rate is not significantly varying with distance from the seafloor, except perhaps for $h < 2$ m at the two bottom-lander T-sensors. This is consistently found for all tidally averaged profiles, of which the spread falls within one order of magnitude, except for $h < 7$ m in which the spread is over two orders of magnitude.

Profiles with tidally averaged turbulent kinetic energy dissipation rate decreasing towards the seafloor outnumber those with increasing values toward the seafloor, by a factor of four. The near-bottom overall mean decrease by half an order of magnitude for $h < 2$ m is found in about half the tidally averaged profiles. The other half shows increasing values towards the seafloor. The eddy diffusivity shows roughly the same tendencies as sketched for dissipation rate, except that the profiles tend to (insignificantly) decrease towards the seafloor, by about a factor of 1.5 over the 56-m vertical range.

*3.4. A glossary of turbulent overturn details*

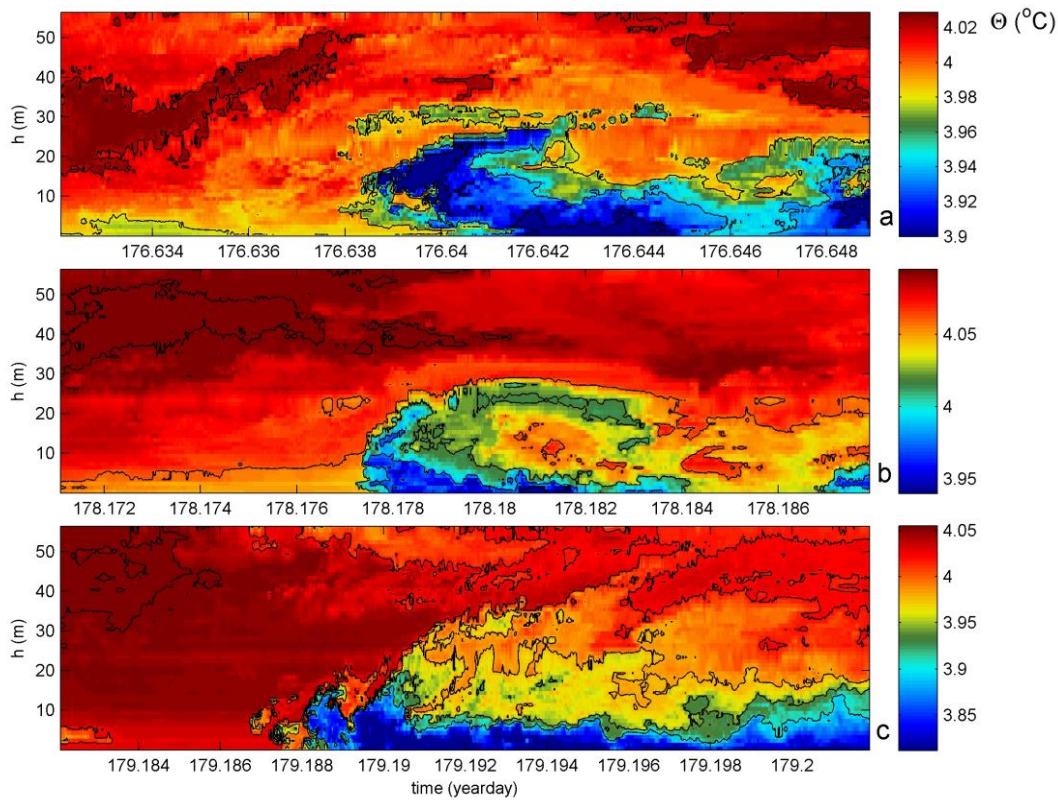

Figure 7

A glossary of upslope propagating and backward breaking frontal bores. In all panels, black isotherms are drawn every 0.04°C and the horizontal axis is at the level of the local seafloor. Note the different colour scales and time ranges. (a) Classic bended front with preceding instabilities guided over it, 1450 s total time-range. (b) Roll-up front with overhanging interior instabilities, 1450 s total time-range. (c) Sharp front with overhanging interior instabilities yielding a turbulence peak, 1710 s total time-range

Even though the 56-m tall T-sensor string did not resolve the tidally dominated internal wave excursions, some of the peaks in turbulent mixing were found reaching (to within 0.4 m from) the sloping seafloor. Fig. 7 demonstrates three examples of upslope propagating frontal bores that induce turbulent mixing lifting the near-seafloor depressed stratification, thereby possibly resuspending loose material several tens of meters above the seafloor. The bores associate with the turning from warming to cooling phase of the internal tide. From the noisy ADCP data it is found that the cores of the bores propagate upward along the thalweg, embedded in a rapid turn within minutes from the northern to the southern canyon sidewall, in cyclonic direction. None of the bore-forms is identical to another, which is probably attributable to variations in precise arrival time, frontal curvature, and local stratification. Their cores extend to about $h = 25$ m, but they have a larger impact via the up- and down sweeping of surrounding waters, as is visible, e.g., in the pathway of warm instabilities. The turbulent kinetic energy dissipation rate locally peaks to a value of $>10^{-6}$ $m^2$ $s^{-3}$, with 56-m and half-hour, Fig. 7-time range, averages of 2.5, 1.7 and $6\times10^{-7}$ $m^2$ $s^{-3}$, for Fig. 7a, 7b, and 7c, respectively.

Other turbulent events occur at some distance from the seafloor, and associate with diapycnal mixing, typically around $h = 20\pm10$ m. Overturns are generally smaller than the bores in Fig. 7, but the examples shown in Fig. 8 display singular waves and Kelvin-Helmholtz instabilities (KHi). Fig. 8a shows a train of interfacial waves, with turbulent overturning underneath the first wave around day 176.325, with progressively shorter time scale and eventual turbulent breaking of the entire wave. The breaking of the short-scale wave occurs underneath an up-going phase of a larger scale wave.

A larger (up-going phase) internal wave can also combine with a train of small-scale KHi around $h = 20$ m (Fig. 8b). As in Fig. 8a, the core around $h = 40$ m under the top of the larger scale wave is turbulent.

In contrast, a larger scale internal wave depression leads to an unusual set of initially vigorous overturns, becoming smaller and less intense with time in arch-like nonlinear waves (Fig. 8c). The asymmetric arch-waves show a shading-like temperature gradient and have no association with commonly modelled KHi roll-up (e.g., Smyth & Moum, 2012).

An example of lesser but still apparent internal wave turbulence and minute KHi around 179.435 is given in Fig. 8d. The interior large-scale instability in the weaker stratified waters above leads the turbulent overturns affecting the local pycnocline.

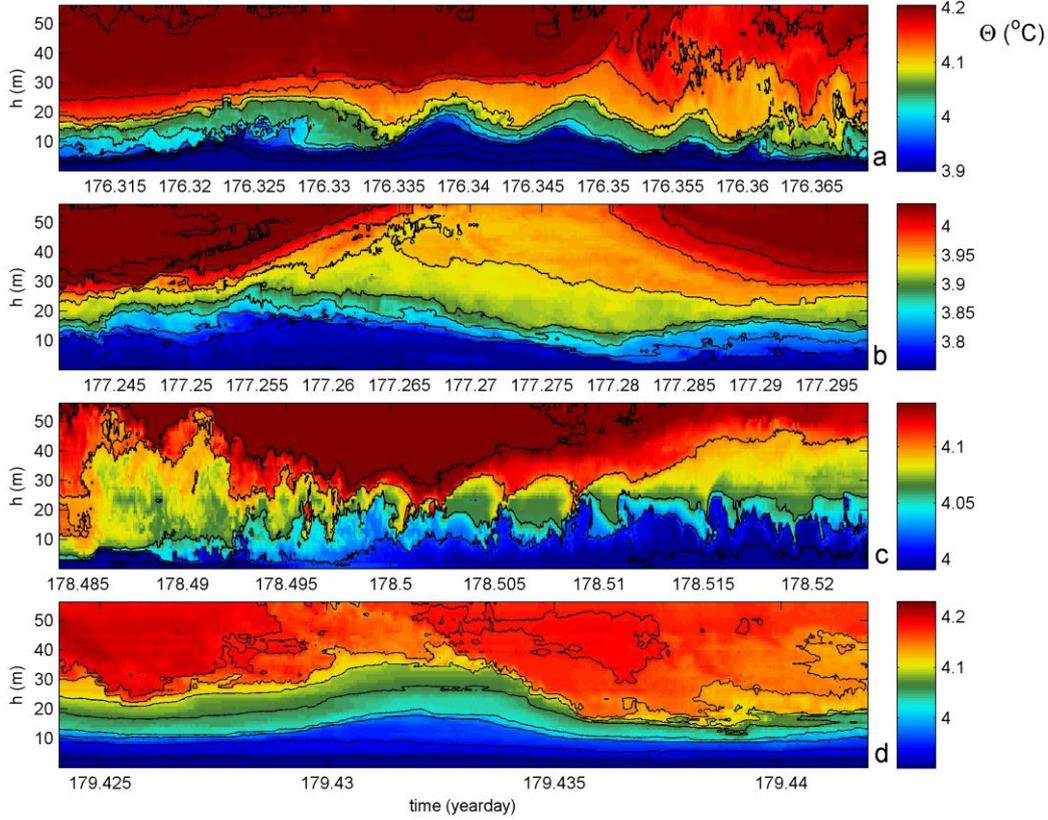

Figure 8

As Fig. 7, but for a glossary of quasi-smooth internal waves and KHi. (a) Small- and large-scale KHi with 3 smooth waves in between, 5000 s total time-range. (b) Large quasi-mode 2 wave, with small-scale KHi, 4900 s total time-range. (c) Small- and large-scale arch-wave like KHi, 3350 s total time-range. (d) Quasi-solitary wave with large turbulent overturns above, 1550 s total time-range

*4. Discussion*

The various forms of nonlinear upslope propagating turbulent bores, singular internal waves, and KHi seem qualitatively somewhat clearer in appearance as laboratory phenomena in near-equatorial deep-ocean observations compared to, e.g., more chaotic mid-latitude GMS observations. However, quantitative tidally averaged turbulence values are in the same order of magnitude for equivalent conditions above topographic slopes that are supercritical for semidiurnal internal waves. Given the near-equatorial dynamics involving the non-approximated full set of equations including the horizontal Coriolis parameter (e.g., Veronis 1963; LeBlond & Mysak, 1978), one expects an influence of (the lack of) rotation on turbulent mixing.

This is because common Earth rotational (Coriolis) effects that dominate large-scale ocean motions through a geostrophic balance become negligible at the equator. Also, general wave breaking in the ocean interior is considered to depend on rotational effects, think of half the open-ocean shear being generated at near-inertial frequencies (Alford, 2003), which is attributed to the short vertical length scales (LeBlond & Mysak 1978). Gregg et al. (2003) used a smooth mainly latitudinal- (and thus inertial frequency-) dependent model to describe their observed 90%-reduction of near-equatorial turbulence kinetic energy dissipation rate in the upper 1000 m of the ocean compared to mid-latitude values. The model they used did not include the non-approximated dynamics, which cause a very sharp decline in near-inertial polarization across $|\varphi| < 2°$ (van Haren, 2005). With the collapse to rectilinear near-inertial motions, sudden increases were observed of large-scale kinetic energy and small-scale density stratification variations. All observations were made from open-ocean data, and none were associated with internal tidal wave motions above underwater topography.

Overall, the here presented turbulence values above near-equatorial underwater topography are equivalent to those from mid-latitudes, to within error. This result seems thus independent of variations in rotational effects and is almost completely attributable to dominant internal tides and their transfer to nonlinear waves above sloping topography. Whilst turbulent mixing associated with the breaking of nonlinear waves does not (primarily) involve shear- rather than convective-overturning and is thus not dependent on dominant near-inertial shear as in the open ocean, it may be informative to compare temperature spectra in the turbulence range from near-equatorial and mid-latitude observations.

In Fig. 9, plotted like the inertial-subrange scaled Fig. 5, near-equatorial temperature variance from moored high-resolution T-sensors is compared with that from GMS at similar heights above the seafloor. Correcting the GMS-spectra with stratification-squared provides almost identical temperature variance at given heights. More importantly, the turbulence-range $N_{max} < \omega <$ roll-off provides very similar spectral slopes at given heights, with most deviating spectrum closest to the seafloor for h = 0.5 m, some convection-turbulence higher up for 2.5 < h < 10-15 m, and dominant shear-turbulence for h > 30 m. The roll-off frequencies have shifted to 1.5-times higher frequencies at GMS, commensurate with the increase in mean buoyancy frequency.

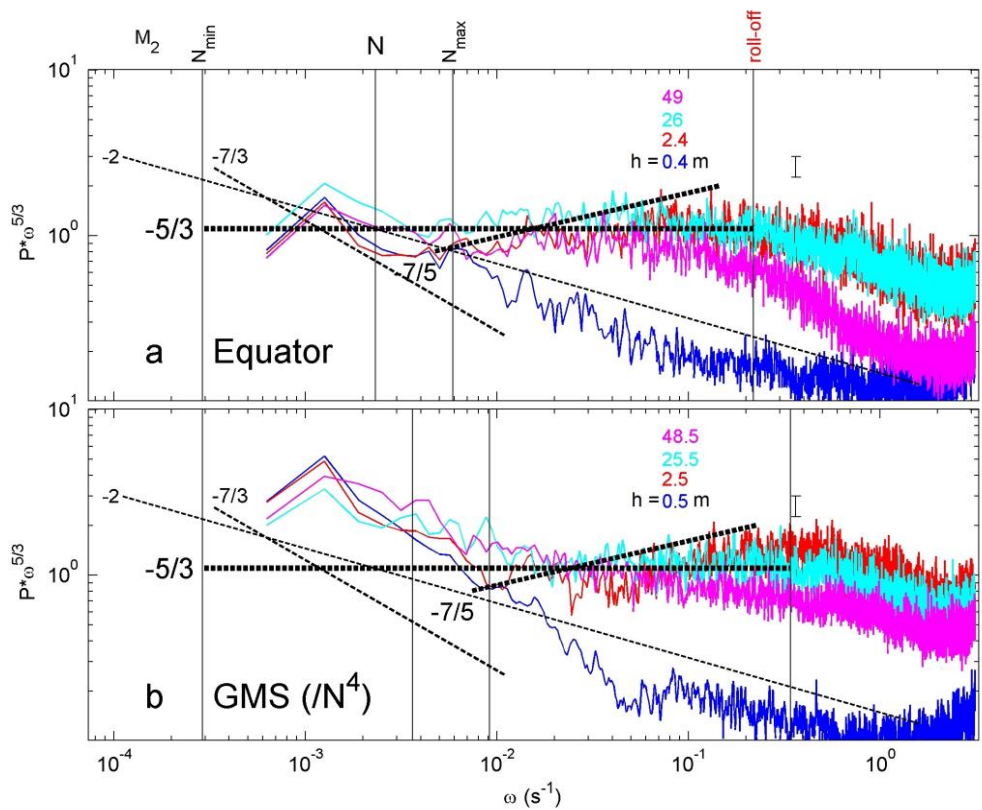

Figure 9

As Fig. 5a, but for a comparison of temperature variance with 6.7-day long T-sensor data from approximately the same heights on a similar bottom-lander mooring some 200 m below the sub-summit of Great Meteor Seamount (GMS) (van Haren and Gostiaux 2012). (a) Equatorial data. (b) GMS data, temperature scaled with the (ratio of) stratification squared

It thus seems that above steep, semi-diurnally supercritical, underwater topography the dominant process is internal (tidal) wave breaking with latitudinal-independent turbulence values. Apart for some less erratic wave deformation near the equator, a consistent build-up

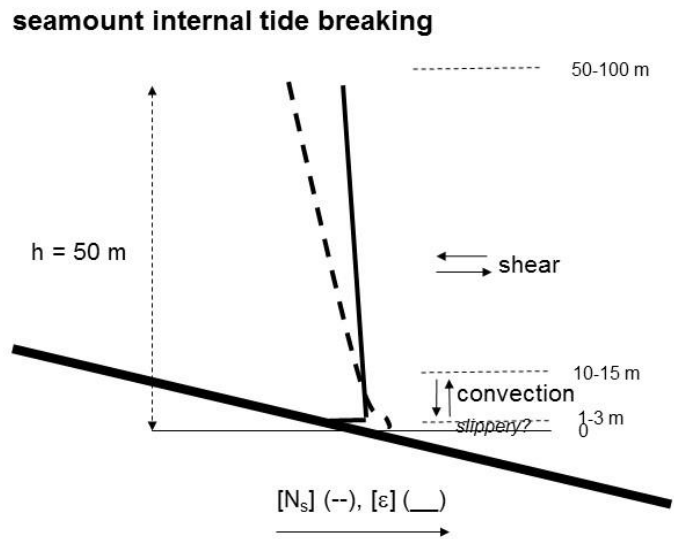

Figure 10

Cartoon of mean profiles for turbulent kinetic energy dissipation rate and buoyancy frequency under internal-tide breaking at a supercritical slope of a near-equatorial seamount, as inferred from moored high-resolution temperature sensors. The turbulence value is barely vertical-dependent, but stratification ($\sim N_s^2$) increases towards the seafloor. A very thin (~2-m) layer above the seafloor shows reduced temperature variance and, if not inflicted by the bottom lander frame, some reduction in turbulence activity. In the next 10 m above, turbulence is dominated by convection as in geothermal heating but here due to wave-breaking, while further up (internal wave) shear dominates turbulence of turbulence-types layering is observed both at mid- and near-equatorial latitudes (Fig. 10). In both areas, the mean semidiurnal tidal flow speed was between 0.12 and 0.15 m s$^{-1}$ (Fig. 3b here and van Haren & Gostiaux 2012). A 1-3-m thick low temperature variance, perhaps slippery, layer is observed underneath an about 10-m thick layer of convection-turbulence dominance in which upslope propagating bores occur and turbulence varies most. Only for h > 15 m, shear-turbulence dominates as familiar as in the open ocean. That is where the short-scale wave- and overturning-forms are observed, albeit without any significant reduction in mean turbulence. The lack of collapse of equatorial internal tidal wave action and associated lack of reduction of turbulent mixing due to wave breaking upon topography may thus be important for deep-ocean stratification.

*Acknowledgements*

I thank the captain and crew of the R/V Pelagia and engineers from NIOZ-MRF for their assistance during mooring preparation, deployment, and recovery. I am indebted to Hokusai-san, possibly with assistance of Katsushika O-Ei, for the perfect imagery including curvature and shading of arch-waves being part of larger scale breaking surface waves, as if depicting internal waves.

*Funding*

NIOZ temperature sensors have been funded in part by NWO, the Netherlands organization for the advancement of science.

*Data availability*

Data supporting the results of this study are available from the author upon reasonable request.

**Declarations**

**Conflict of interest** The author declares no competing interests.

*Appendix A*

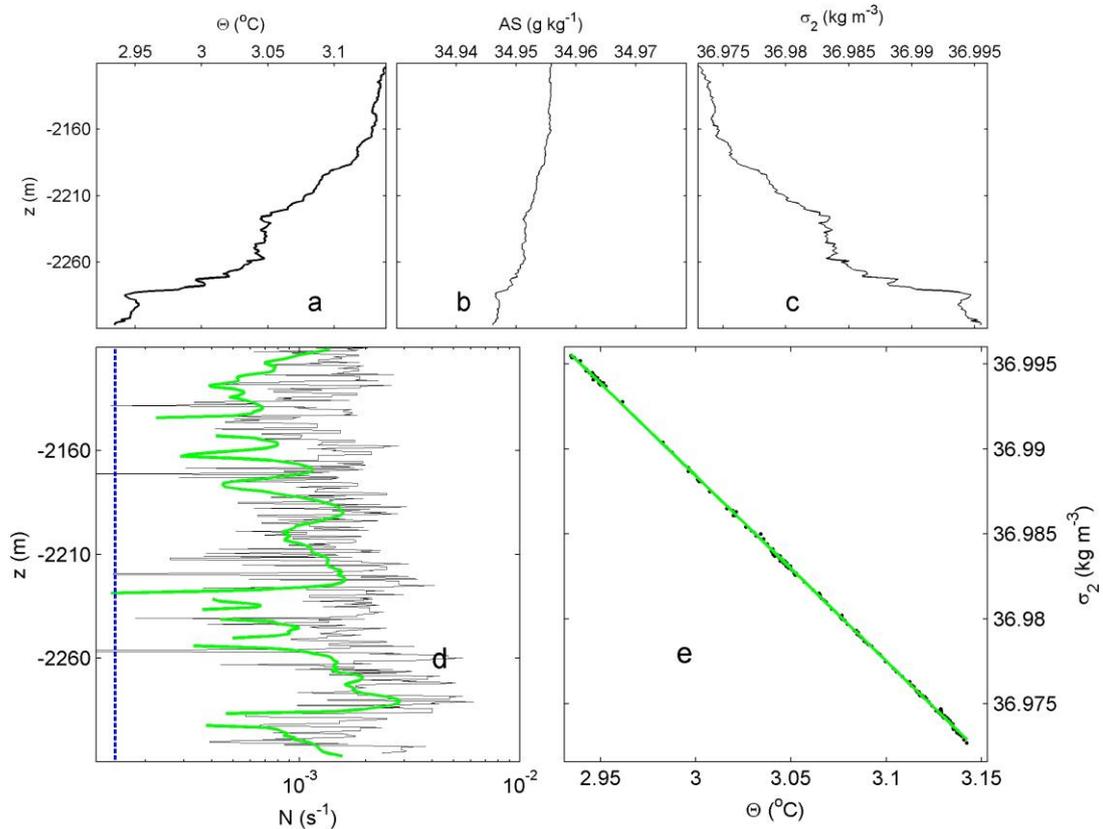

Figure 11

As Fig. 2, but for lower 200-m part of a CTD-profile, down to 5 m above the seafloor

Commonly, shipborne CTD profiles are made as close as possible near the mooring containing high-resolution T-sensors. This is to achieve best possible local information on the temperature profile(s) via recently calibrated instrumentation and on the temperature-density relationship. During the present cruise, the nearest profiles were obtained some 3 km north of the T-sensor mooring, where the water depth was about 600 m more than at the mooring site. This hampers a direct comparison, as the stratification is commonly higher closer to the sea surface, whilst ocean-interior stratification away from topography is less turbulent and may show partial salinity-compensated temperature inversions that are not (yet) mixed away. Therefore, two portions of a CTD-profile are provided in this paper. In Fig. 2, profiles are given from around the depth of the mooring, hence of interior waters. In Fig. 11, the associated profiles are given from the lower 300 m above the seafloor at the CTD-station.

This figure shows profiles that are more ragged, turbulent, and yet provide an overall tighter temperature-density relationship, compared to Fig. 2. While the overall value for buoyancy frequency is larger in Fig. 2, it is seen to increase its value with increasing depth in Fig. 11, which is in line with the observations from T-sensors in Fig. 6a.